\documentclass[pra, twocolumn, showpacs, showkeys, eqsecnum]{revtex4}
\usepackage{graphics,times}

\newcommand {\psig}{\Sigma_}
\newcommand {\pxi}{\Xi_}

\newcommand {\avecsig}{\langle \vec{\Sigma} \rangle}

\begin{document}
\title{A hazard of open quantum dynamics: Markov approximations encounter map
domains}
\author{Thomas F. Jordan}
\email[email: ]{tjordan@d.umn.edu}
\affiliation{Physics Department, University of Minnesota, Duluth, Minnesota
55812}
\author{Anil Shaji}
\email[email: ]{shaji@unm.edu}
\affiliation{Department of Physics and Astronomy, The University of New Mexico,
800 Yale Boulevard NE, Albuquerque, New Mexico 87131}
\author{E. C. G. Sudarshan}
\email[email: ]{sudarshan@physics.utexas.edu}
\affiliation{Center for Statistical Mechanics, The University of Texas at
Austin, 1 University Station C1609, Austin, Texas 78712}  

\begin{abstract}
A Markov approximation in open quantum dynamics can give unphysical results
when a map acts on a state that is not in its domain. This is examined here in a
simple example, an open quantum dynamics for one qubit in a system of two
interacting qubits, for which the map domains have been described quite
completely. A time interval is split into two parts and the map from the exact
dynamics for the entire interval is replaced by the conjunction of that same map
for both parts. If there is any correlation between the two qubits, unphysical
results can appear as soon as the map conjunction is used, even for
infinitesimal times. If the map is repeated an unlimited number of times, every
state is at risk of being taken outside the bounds of physical meaning.
Treatment by slippage of initial conditions is discussed. 

\end{abstract}


\maketitle

\section{Introduction}\label{one}

A Markov approximation assumes that the way a state changes in time does not
depend on the time when the change begins. In open quantum dynamics, this
assumption does not hold when there are correlations between the subsystem being
considered and the rest of the larger system, because the changes of states in
time are described by maps that depend on the correlations, and the correlations
change in time. The maps generally are not completely positive and act on
limited domains \cite{jordan04a,jordan06a}. Maps for changes that begin at
different times can have different domains. When a Markov approximation requires
the same map to act at different times, it may require the map to act on a state
that is not in its domain. This can give unphysical results. A density matrix
for a state can be mapped to a matrix that is not a density matrix and can not
represent a physical state.

We look at this here in a simple example, an open quantum dynamics for one qubit
in a system of two interacting qubits, for which the map domains have been
described quite completely \cite{jordan04a,jordan06a}. We consider only one
particular hazard that Markov approximations encounter in open quantum dynamics.
We examine only the Markov property that the same map is used at different
times. We use maps from the exact dynamics. They are not approximations. We just
use them out of place. We split a time interval into two parts and replace the
map for the entire interval with the conjunction of that same map for both
parts. If there is any correlation at all between the two qubits, unphysical
results can appear as soon as the map conjunction is used, even for
infinitesimal times. If the map is repeated an unlimited number of times, every
state is at risk of being taken outside the bounds of physical meaning. In this
simple example, everything that happens can be seen clearly. 

A Markov approximation is not expected to work very well for this example
because the reservoir is no bigger than the system and changes as fast as the
system does. There can be large correlations between the system and the
reservoir and they can change just as fast. The exact dynamics of the system
does not depend on time as a semigroup. Our interest is not in a failure of
accuracy of a Markov approximation. It is in the reason for unphysical results.
We show that unphysical results can occur simply because a map is made to act on
a state that is not in its limited domain. Our focus is on map domains. We use
this two-qubit example because it is the only example we have where the domains
are known. 

Markov approximations in open quantum dynamics encounter many hazards. Knowing
the nature of this one particular hazard, seen clearly in this example, may make
it easier to navigate around it in other settings. Separating it from other
hazards may make it possible to view them more clearly.

To be able to see what happens, we use maps of mean values to describe both the
open dynamics of the single qubit and the relevant parts of the full dynamics of
the two qubits. This keeps us within sight of established navigation marks. The
map domains are described in terms of mean values. They depend on correlation
mean values that make the maps change with time. These marks are not visible
when you work with a master equation or use an operator sum form to describe the
maps.

The importance of correlations has been long appreciated
\cite{haake83a,haake85a}. The role of map domains has not been described. One
measure used to deal with hazards of unphysical results is to adjust or ``slip"
the initial conditions
\cite{suarez92a,gaspard99a,wilkie01a,benatti03a,benatti06a}. This is discussed,
in the light of the example, in Section IV. First the example is set out, the
exact dynamics in Section II, and the map conjunctions in Section III.

\section{Exact dynamics}\label{two}

We consider two qubits, one described by Pauli matrices $\Sigma_1$, $\Sigma_2$,
$\Sigma_3$ and the other by Pauli matrices $\Xi_1$, $\Xi _2$, $\Xi_3$, and
consider the dynamics generated by the Hamiltonian
\begin{equation}
\label{Ham}
H = \frac{1}{2}\psig 3 \pxi 1 .
\end{equation}
We focus on the open dynamics of the $\Sigma $ qubit described by maps of the
mean values $\langle \Sigma_1 \rangle $, $\langle \Sigma_2 \rangle $, $\langle
\Sigma_3 \rangle $ at time $0$ to
\begin{eqnarray}
\label{sigt}
\langle \Sigma_1 \rangle (t) & = & \langle \Sigma_1 \rangle \cos t - \langle
\Sigma_2 \Xi_1 \rangle \sin t \nonumber \\
\langle \Sigma_2 \rangle (t) & = & \langle \Sigma_2 \rangle \cos t + \langle
\Sigma_1 \Xi_1\rangle \sin t \nonumber \\
\langle \Sigma_3 \rangle (t) & = & \langle \Sigma_3 \rangle 
\end{eqnarray}
at times $t$. We write $\avecsig$ for the vector with components $\langle
\Sigma_1 \rangle $, $\langle \Sigma_2 \rangle $, $\langle \Sigma_3 \rangle $, or
for the set of those three components, and write $\avecsig (t)$ for $\langle
\Sigma_1 \rangle (t)$, $\langle \Sigma_2 \rangle (t)$, $\langle \Sigma_3 \rangle
(t)$. For each $t$ and each $\langle \Sigma_2 \Xi_1 \rangle $ and $\langle
\Sigma_1 \Xi_1 \rangle $, there is a map from states at time $0$, described by
mean values $\avecsig $, to the states at time $t$ described by the mean values
$\avecsig (t)$. Different $\langle \Sigma_2 \Xi_1 \rangle $ and $\langle
\Sigma_1 \Xi_1 \rangle $ give different maps.

When there are correlations between the two qubits, these maps generally are not
completely positive and act in limited domains. Each map is made to be used for
a particular set of states described by a particular set of $\avecsig$, which we
call the compatibility domain. It is the set of $\avecsig$ that are compatible
with the $\langle \Sigma_2 \Xi_1 \rangle $ and $\langle \Sigma_1 \Xi_1 \rangle $
in describing a possible initial state for the two qubits. In a larger domain,
which we call the positivity domain, the map takes every positive matrix to a
positive matrix. The positivity domain is the set of $\avecsig$ for which
$|\avecsig (t)| \leq 1$. We have described these domains quite completely for
the maps we consider here \cite{jordan04a,jordan06a}. For these maps, the
compatibility domain is the intersection of all the positivity domains for
different $t$ for the same values of $\langle \Sigma_2 \Xi_1 \rangle $ and
$\langle \Sigma_1 \Xi_1 \rangle $; this is not generally true
\cite{jordan04a,jordan06a}.

Here, with no loss of generality, we let $\langle \Sigma_2 \Xi_1 \rangle $ be
zero and $\langle \Sigma_1 \Xi_1 \rangle $ positive. Then the maps depend on
just the one parameter $\langle \Sigma_1 \Xi_1 \rangle $. It changes in time to
\begin{equation}
\label{sigxit}
\langle \Sigma_1 \Xi_1 \rangle (t) =  \langle \Sigma_1 \Xi_1 \rangle \cos t -
\langle \Sigma_2 \rangle \sin t.  
\end{equation}

\section{Map conjunctions}\label{three}

Let $\avecsig (t|s)$ be the replacement for $\avecsig (t+s)$ obtained by going
from time $0$ to time $t$ with the map established at time $0$ with $\langle
\Sigma_1 \Xi_1 \rangle $ and then from time $t$ to time $t+s$ with the same map,
with $\langle \Sigma_1 \Xi_1 \rangle $ unchanged. We call this a map
conjunction. We use it to examine one property of Markov approximations that can
give unphysical results: application of the same map at different times. This
replacement is not a Markov approximation itself. The semigroup property is used
only at the point where the maps are joined. The maps are from the exact
dynamics. They are not approximations. They are just being used out of place. 

We look at this map conjunction for initial states where $\langle \Sigma_1
\rangle $ and $\langle \Sigma_3 \rangle $ are zero. Then $\langle \Sigma_1
\rangle (t)$, $\langle \Sigma_3 \rangle (t)$ and $\langle \Sigma_1 \rangle
(t|s)$, $\langle \Sigma_3 \rangle (t|s)$ are all zero, and
\begin{eqnarray}
\label{sigts}
\langle \Sigma_2 \rangle (t|s) & = & \langle \Sigma_2 \rangle (t)\cos s +
\langle \Sigma_1 \Xi_1 \rangle \sin s \nonumber \\
 & = & \langle \Sigma_2 \rangle \cos t \cos s \nonumber \\
&& \hspace{5 mm} + \langle \Sigma_1 \Xi_1\rangle (\sin t \cos s + \sin s ).
\end{eqnarray}

When $\langle \Sigma_1 \rangle $ and $\langle \Sigma_3 \rangle $ are zero,
$\avecsig$ is in the compatibility domain for $\langle \Sigma_1 \Xi_1 \rangle $
if \cite{jordan04a,jordan06a}
\begin{equation}
\label{domain}
\langle \Sigma_2 \rangle^2 +  \langle \Sigma_1 \Xi_1 \rangle^2 \leq  1.  
\end{equation}
To push to the limit, we look at a point on the edge of the compatibility domain
and let
\begin{equation}
\label{qpoint} 
\langle \Sigma_1 \Xi_1 \rangle = \sin q, \quad \quad \langle \Sigma_2 \rangle =
\cos q,  
\end{equation}
both positive. In particular, we look at what happens when $t$ is $q$. We have
\begin{eqnarray}
\label{atq}
\langle \Sigma_2 \rangle (q) & = & 1 \nonumber \\
\langle \Sigma_1 \Xi_1 \rangle (q) & = & 0 \nonumber \\
\langle \Sigma_2 \rangle (q+s) & = & \cos s \nonumber \\
\langle \Sigma_2 \rangle (q|s) & = & \cos s + \sin q \sin s.
\end{eqnarray}
When $s$ is $0$, both $\langle \Sigma_2 \rangle (q+s)$ and $\langle \Sigma_2
\rangle (q|s)$ are $1$. Then, as $s$ increases, $\langle \Sigma_2 \rangle (q+s)$
goes down, but $\langle \Sigma_2 \rangle (q|s)$ goes up; when $s$ is $0$, the
slope $d\langle \Sigma_2 \rangle (q|s)/ds$ is $\sin q$, which we are assuming is
positive. We have the physically impossible result that $\langle \Sigma_2
\rangle (q|s)$ becomes larger than $1$; it can not be a mean value $\langle
\Sigma_2 \rangle $ for a physical state. This happens for any $\langle \Sigma_1
\Xi_1 \rangle $ that is not zero. It happens for infinitesimal $s$. If there is
any correlation at all between the two qubits, unphysical results appear as soon
as the map conjunction is used.

The unphysical results occur when the map acts on a state that is not in its
domain. From Eqs.(\ref{domain}), we see that when $t$ is $q$ and $\langle
\Sigma_2 \rangle (t)$ is $1$, the state is not in the compatibility domain of
the map. It is in the compatibility domain for the map that could be established
at time $q$ to follow the exact dynamics, because $\langle \Sigma_1 \Xi_1
\rangle (t)$ is $0$ when $t$ is $q$. 

Now we look at what can happen with more repetitions of the map. Let $\avecsig
(t|s_1 |s_2 )$ be the replacement for $\avecsig (t+s_1 +s_2)$ obtained by going
through three steps in time from $0$ to $t$ to $t+s_1$ to $t+s_1 +s_2 $ using
the same map established at time $0$, with the same $\langle \Sigma_1 \Xi_1
\rangle $, for each step. In general, let $\avecsig (t|s_1 | \, ... \, |s_n )$
be the replacement for \linebreak $\avecsig (t+s_1 + ... +s_n )$ obtained by going through
$n+1$ steps in time from $0$ to $t$ to $t+s_1 $ ... to $t+s_1 + ... +s_n $ using
that same map for each step. Now we want to look at states deep inside the
compatibility domain, not on the edge, so we do not assume Eqs.~(\ref{qpoint}),
and we do not assume that $t$ is $q$. We do still assume that $\langle \Sigma_1
\rangle $ and $\langle \Sigma_3 \rangle $ are zero. From Eqs.~(\ref{sigt}), we
can see that as $t$ varies, the magnitude of $\langle \Sigma_2 \rangle (t)$ is
maximum when
\begin{equation}
\label{maxt}
[\langle \Sigma_2 \rangle (t)]^2 = \langle \Sigma_2 \rangle^2 +  \langle
\Sigma_1 \Xi_1 \rangle^2 .  
\end{equation}
From this and the first line of Eq.~(\ref{sigts}), we can see that as $t$ and
$s$ vary, the magnitude of $\langle \Sigma_2 \rangle (t|s)$ is maximum when
\begin{eqnarray}
\label{maxts}
[\langle \Sigma_2 \rangle (t|s)]^2 & = & [\langle \Sigma_2 \rangle (t)]^2 +
\langle \Sigma_1 \Xi_1 \rangle^2 \nonumber \\
 & = & \langle \Sigma_2 \rangle^2 +  2\langle \Sigma_1 \Xi_1 \rangle^2 .  
\end{eqnarray}
We can see similarly that as $t$, $s_1 $ and $s_2$ vary, the magnitude of
$\langle \Sigma_2 \rangle (t|s_1 |s_2 )$ is largest when
\begin{eqnarray}
\label{maxt12}
[\langle \Sigma_2 \rangle (t|s_1 |s_2 )]^2 & = & [\langle \Sigma_2 \rangle
(t|s_1 )]^2 + \langle \Sigma_1 \Xi_1 \rangle^2 \nonumber \\
 & = & \langle \Sigma_2 \rangle^2 +  3\langle \Sigma_1 \Xi_1 \rangle^2 ,  
\end{eqnarray}
and as $t$, $s_1 $, ... and $s_n$ vary, the magnitude of $\langle \Sigma_2
\rangle (t|s_1 |\, ... \, |s_n )$ is largest when
\begin{eqnarray}
\label{maxt1n}
[\langle \Sigma_2 \rangle (t|s_1 |\, ... \, |s_n )]^2 & = & [\langle \Sigma_2
\rangle (t|s_1 |\, ... \, |s_{n-1} )]^2  + \langle \Sigma_1 \Xi_1 \rangle^2
\nonumber \\
 & = & \langle \Sigma_2 \rangle^2 +  (n+1)\langle \Sigma_1 \Xi_1 \rangle^2 .  
\end{eqnarray}
As $n$ increases, the magnitude of $\langle \Sigma_2 \rangle (t|s_1 |\, ... \,
|s_n )$ can grow with each step and eventually become larger than $1$. This can
happen no matter how small $\langle \Sigma_2 \rangle $ and $\langle \Sigma_1
\Xi_1 \rangle $ may be, as long as $\langle \Sigma_1 \Xi_1 \rangle $ is not
zero. If there is any correlation at all between the two qubits, every state of
the $\Sigma $ qubit is at risk of being taken outside the bounds of physical
meaning.

Again, unphysical results appear when the map acts on a state that is not in its
domain. From Eqs.~(\ref{domain}) and (\ref{maxt1n}), we can see that when the
magnitude of $\langle \Sigma_2 \rangle (t|s_1 |\ldots |s_n )$ is larger than
$1$, the predecessor $\langle \Sigma_2 \rangle (t|s_1 |\ldots |s_{n-1} )$ is
not a mean value $\langle \Sigma_2 \rangle $ for a state in the compatibility
domain of the map. Repetitions of the map can move a state out of the
compatibility domain from deep inside it.

\section{Slipping initial conditions}\label{four}

One measure used to deal with hazards of unphysical results is to adjust or
"slip" the initial conditions, or restrict what they can be
\cite{suarez92a,gaspard99a,wilkie01a,benatti03a,benatti06a}. The maps are kept
the same. In our example, we see that unphysical results can indeed be
eliminated by changing or restricting the initial value $\langle \Sigma_2
\rangle $. From Eqs.~(\ref{domain}), (\ref{maxt}) and (\ref{maxts}), we see that
the magnitude of $\langle \Sigma_2 \rangle (t|s)$ remains smaller than $1$ for
all $t$ and $s$ if
\begin{equation}
\label{domainatt}
\langle \Sigma_2 \rangle^2 +  2\langle \Sigma_1 \Xi_1 \rangle^2 \leq  1,  
\end{equation}
which means that the state represented by $\avecsig (t)$ is in the compatibility
domain. This is more restrictive than the requirement (\ref{domain}) that the
state represented by $\avecsig $ is in the compatibility domain, but it may be
an acceptable restriction when the correlation between the two qubits described
by $\langle \Sigma_1 \Xi_1 \rangle $ is small. From Eq.(\ref{maxt1n}), we see
that
\begin{equation}
\label{domainatt1}
\langle \Sigma_2 \rangle^2 +  (n+1)\langle \Sigma_1 \Xi_1 \rangle^2 \leq  1  
\end{equation}
would be needed to prevent unphysical results when $n$ repetitions of the map
are arranged to occur at times when they do maximum damage. Such an arrangement
is not likely to occur in a well constructed model, but knowing how it could
occur allows care to be taken to see that it does not.

In cases where unphysical results have been eliminated by restricting initial
conditions, the maps of the remaining initial states have been observed to
produce entanglement with another system that is separate from the dynamics
\cite{benatti03a,benatti06a}. This is not necessarily a symptom of pathology.
Such entanglement can be produced by the maps that describe the correct exact
 open quantum dynamics of a subsystem when there are correlations between that
subsystem and the rest of the larger system involved in the dynamics
\cite{jordan07a}.

\acknowledgements

One of the authors (A.~S.) acknowledges financial support from the Office of
Naval Research Grant No. N00014-07-1-0304.

\bibliography{masterequation}
\end{document}